\documentclass[superscriptaddress,twocolumn,showpacs,preprintnumbers,amsmath,amssymb,prl]{revtex4}

\usepackage{graphicx,psfig,rotating,color,psfrag,amssymb,epsfig}
\usepackage{dcolumn}
\usepackage{bm}

\newcommand{\bfr}{{\bf r}}

\begin{document}


\title{Observations on Sound Propagation in Rapidly Rotating Bose-Einstein Condensates}

\author{T.~P. Simula}
\affiliation{Department of Physics, University of Otago, Dunedin, New Zealand}
\author{P. Engels}
\affiliation{JILA, National Institute of Standards and Technology and\\
Department of Physics, University of Colorado, Boulder, Colorado 80309-0440, USA}
\author{I. Coddington}
\affiliation{JILA, National Institute of Standards and Technology and\\
Department of Physics, University of Colorado, Boulder, Colorado 80309-0440, USA}
\author{V. Schweikhard}
\affiliation{JILA, National Institute of Standards and Technology and\\
Department of Physics, University of Colorado, Boulder, Colorado 80309-0440, USA}
\author{E.~A. Cornell}
\altaffiliation{Quantum Physics Division, National Institute of Standards and Technology.}
\affiliation{JILA, National Institute of Standards and Technology and\\
Department of Physics, University of Colorado, Boulder, Colorado 80309-0440, USA}
\author{R.~J. Ballagh}
\affiliation{Department of Physics, University of Otago, Dunedin, New Zealand}

\date{\today}
\begin{abstract}
Repulsive laser potential pulses applied to vortex lattices of rapidly rotating Bose-Einstein condensates create propagating density waves which we have observed experimentally and modeled computationally to high accuracy. We have observed a rich variety of dynamical phenomena ranging from interference effects and shock-wave formation to anisotropic sound propagation.
\end{abstract}

\maketitle

\emph{Introduction.---}Superfluids can be described in terms of an order parameter associated with the superfluid phase of the system. In the case of dilute Bose-Einstein condensates a natural choice for the order parameter is the macroscopic wavefunction, whose squared modulus yields the particle density of the condensate. It follows that the superflow, given by the gradient of the phase of the condensate, must be irrotational and therefore rotation of a superfluid requires the presence of quantized vortices with cores of vanishing density \cite{DonnellyBook}. The minimum energy configuration of a rotating superfluid is a triangular Abrikosov lattice of vortices as was originally established in the context of He-II. This fundamental superfluid behavior has also been witnessed in dilute Bose-Einstein condensates \cite{Madison2000a,Abo-Shaeer2001a,Haljan2001a,Hodby2002a}.

The structure and dynamics of single vortices in dilute Bose-Einstein condensates have been extensively studied \cite{Fetterreview}. The presence of multiple vortices radically changes the spectral properties of the condensate while a relative motion of the interacting vortices enriches the superfluid dynamics. Recently, collective behavior of vortex lattices has attracted much theoretical and experimental interest. Oscillations of the giant vortex core area \cite{Engels2003a,Simula2004a,Fetter2003a}, excitation of vibrational Tkachenko modes \cite{Engels2003a,Coddington2003a,Simula2004a,Tkachenko1966a,Baym2004a,Mizushima2004a,Baksmaty2004a}, and the structure of the condensate in the lowest Landau level \cite{Ho2001a,Schweikhard2004a,Baym2004b} have been observed by changing the moment of inertia of the rotating condensate by a selective atom removal. The tilting mode of a vortex array has been observed by applying a method analogous to the driving of the scissors mode of a non-rotating condensate \cite{Smith2004a}.

In dilute Bose-Einstein condensates, compressional sound waves correspond to modulations in the condensate density. Propagation of such sound waves were observed in harmonically trapped ground state condensates \cite{Ketterle1997a}, and in two-component condensates density defects were seen to deform into vortex rings and dark solitons \cite{Anderson2001a,Dutton2001a}. The effect of an optical lattice on sound propagation has also attracted attention recently \cite{Menotti2004a}. Since the condensate density vanishes at vortex cores, the rotation of the condensate is expected to have notable effects on the sound propagation. In this Letter we study both experimentally and theoretically the response of a rapidly rotating Bose-Einstein condensate to a variety of external repulsive laser pulses. We compare our experimental observations and \emph{a priori} time-dependent numerical simulations, and show how the topological constraints set by the vortices result into exotic non-equilibrium dynamics such as anisotropically propagating sound waves.

\begin{figure}[!t]
\includegraphics[width=86mm]{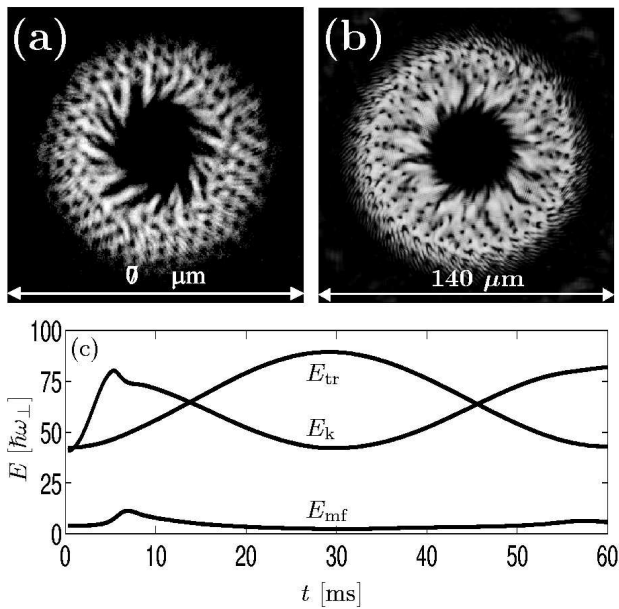}
\caption{Experimental (a) and theoretical (b,c) response of a
vortex lattice to a localized 5 ms long repulsive laser pulse.
Beam parameters are $P=0.64$ mW and $d=19\;\mu$m. The experimental
image is taken after 45 ms anti-trapped expansion, whereas the 3D computational image is taken after 14 ms of in-trap evolution. }
\label{Narrowbeam}
\end{figure}

\emph{Experiment and theory.---}The experimental results presented in this paper are obtained by
starting with rapidly rotating $^{87}$Rb condensates held in a
cylindrically symmetric trap with trapping frequencies
$\{\omega_\perp,\omega_{z}\}=2\pi\times\{8.3,5.3\}$ Hz. The condensates
typically contain $N=3.5\times10^{6}$ atoms and are rotating around
the $z$-axis with a rotation rate of $0.95\;\omega_\perp$. To
create blastwaves, a far off-resonant laser beam, centered on the
condensate, is sent through it along the $z$-axis. This beam
has a wavelength of 660~nm, which is far detuned from the Rb D2
line at 780~nm. It creates a dipole potential given by
\begin{equation}
V_{\rm dip}(\bfr)= k_{\rm B} \times 129 \mu {\rm K} \left(\frac{1 \mu
{\rm m}}{d}\right)^2 \left(\frac{P}{1 {\rm mW}}\right) e^{-2r^2/d^2}
\label{dippot}
\end{equation}
such that for a beam with power $P=0.64$ mW and waist $d=19\;\mu$m as
used for Fig.~\ref{Narrowbeam}, the atoms at the center of the
laser beam experience a repulsive potential with characteristic
frequency of $2 \pi \times 78$ Hz. The potential height in this case
surpasses the chemical potential of the condensate by a factor 39.
After the pulse, the condensate is expanded by an anti-trapping
geometry as described in \cite{Coddington2004a}, and imaged along
the axis of rotation.


The experiment is simulated in the mean-field formalism by calculating the full 3D evolution of the condensate wavefunction with the time-dependent Gross-Pitaevskii equation. We adopt $\hslash\omega_\perp$ for the energy scale and $a_0=\sqrt{\hslash/2m\omega_\perp}$ for the spatial length scale. From the computational point of view, in the absence of the laser field, the condensate is fully characterized by the dimensionless values for the nonlinearity constant $C=gN/\hslash\omega_\perp a_0^3$, and the trap anisotropy parameter $\lambda=\omega_z/\omega_\perp$, when the wavefunction $\psi$ is normalized to unity. Above, $g$ is the usual $s$-wave coupling constant. Modeling this experiment we choose $C= 1.6 \times 10^5$ and $\lambda=0.64$.

\begin{figure}[!t]
\includegraphics[width=86mm]{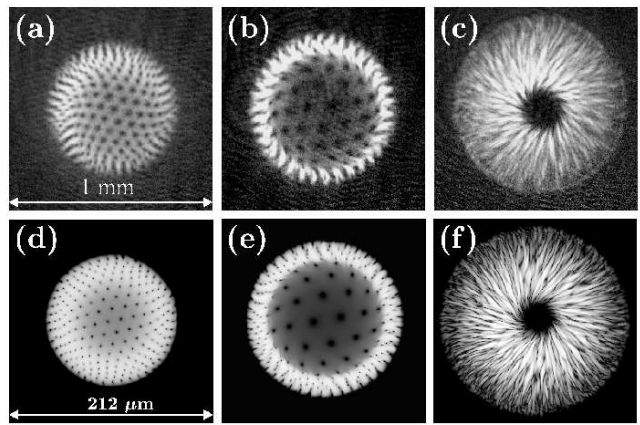}
\caption{Response of a vortex lattice to wide repulsive laser pulses. Upper row displays experimental pictures and the corresponding frames below them are from 3D numerical simulations. The pulse durations are 5 ms (a,d), 10 ms (b,e), and 30 ms (c,f). The experimental (computational) pictures are taken after 45~ms of anti-trapped expansion (respectively, $(12,16$, and $22)$ ms in-trap evolution) after the end of the pulse.}
\label{Largebeam}
\end{figure}

\emph{Wide pulse.---}Figure \ref{Largebeam} shows experimental observations (upper row) and their theoretical counterparts (lower row) of the integrated condensate density, for the case where a laser pulse of power $P=4.5$ mW and waist $d=60\;\mu$m (roughly the radius of the condensate) is applied for varying lengths of time. The pulse durations are 5 ms (a,d), 10 ms (b,e), and 30 ms (c,f). For the shortest pulse (a,d), mainly the collective modes such as the Tkachenko waves and longitudinal breathing modes are excited. When the duration of the pulse is doubled (b,e), the central part of the condensate is repelled to form a narrow outer band of high density where the condensate is channeled between the vortex planes forming superfluid flares. The vortices are compressed both in core size and spacing there. In contrast, vortices in the central region experience an outward movement and expand in both core size and spacing due to the strongly reduced condensate density. It is notable that the number of particles per vortex in the centermost region remain practically unaltered from its equilibrium value. For even longer pulse duration (c,f) a five-fold increase in the trapping energy of the condensate takes place. When this excess energy is converted into kinetics of the condensate, strong radial currents result and the stripes seen in (c,f) emerge due to the fact that the vortex cores constitute forbidden positions for the superfluid flow. The outer higher density halo at the condensate horizon is a remnant of the earlier state when the condensate assumed a shape of a thin cylinder. 

\begin{figure}[!t]
\includegraphics[width=86mm]{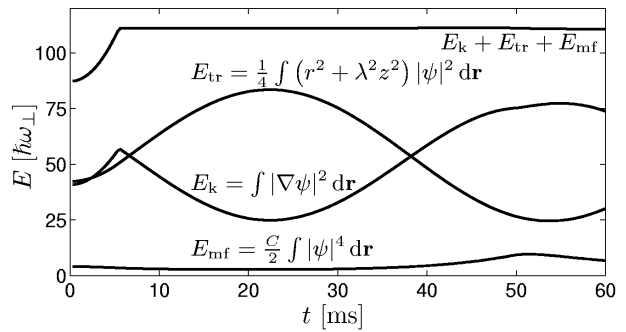}
\caption{Kinetic $E_{\rm k}$, trap $E_{\rm tr}$, and mean-field $E_{\rm mf}$ energies and their sum as functions of time. The data is obtained from the same simulation as Fig.~\ref{Largebeam}(d).}
\label{Egyplot1}
\end{figure}

The physics seen in Fig.~\ref{Largebeam} is conveniently analysed using Fig.~\ref{Egyplot1} where different components of energy are plotted as functions of time corresponding to the simulation in Fig.~\ref{Largebeam}(d). All of the three cases, Fig.~\ref{Largebeam}(d)-(f), exhibit similar charactristics in terms of absorbtion of energy from the beam followed by a subsequent oscillatory exchange of kinetic and potential energies at the characteristic radial compressional mode frequency of $2.0 \omega_\perp$ \cite{Cozzini2003a}. However, the amount of energy absorbed by the condensate from the beam depends on the pulse duration, resulting in the qualitatively different features between the tree different cases in Fig.~\ref{Largebeam}.

The anti-trapping stage is not modeled in the simulations presented due to the limitation of computational resources. This may slightly alter the apparent size of the vortices due to the axial evolution as explained in \cite{Coddington2004a}. However, as is shown in Fig.~\ref{Egyplot1}, the mean-field energy amounts to only a few percent of the total energy and thus the dynamical evolution is mainly governed by the interplay between kinetic and potential energies, which explains the similarity between the in-trap-and anti-trapped results, apart from the scaling in space and time.

\begin{figure}[!t]
\includegraphics[width=86mm]{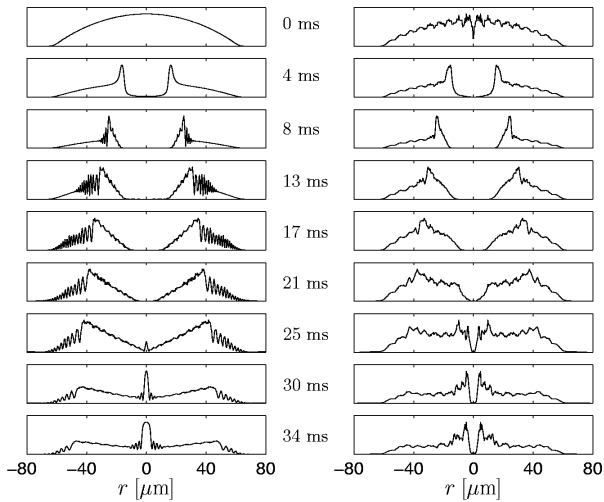}
\caption{Simulations (3D) on sound propagation in a ground state (lhs) and in a vortex lattice (rhs). The ordinates are normalized according to the temporal peak density.}
\label{GroundvsLattice}
\end{figure}

\emph{Narrow pulse.---}When applying laser pulses with narrow beam waist (with respect to the width of the condensate), travelling density perturbations are created. Strong applied power produces shock-waves, fringes and stripes as highlighted in Fig.~\ref{Narrowbeam}, whereas if a weak pulse is used, smoothly propagating sound waves are formed as demonstrated at the end of the paper. The physics in Fig.~\ref{Narrowbeam} is again characterised by the energy development. At early times, the kinetic energy is greatly increased as the strong laser potential pushes atoms at supersonic speeds away from the beam focus thereby generating a steep wavefront that rapidly propagates radially while experiencing a twist under the influence of the Coriolis effect. The formation of the shock front at the edge of the beam is manifested by a sudden increase in the mean-field energy shortly after the end of the pulse as shown in Fig.~\ref{Narrowbeam}(c). The average initial velocity $\langle v\rangle$ of the repelled atoms may be crudely estimated by distributing the kinetic energy absorbed from the beam $\Delta E_k$ over the particles within the beam volume $V=2\pi d^2 z_{\rm TF}$. The result is $ \langle v\rangle=c_s\sqrt{2C\hslash\omega_\perp\Delta E_k a_0^3/V\mu^2}$, where $c_s=\sqrt{\mu/m}$ is the speed of sound. This provides a prediction for the shock formation and for the case in Fig.~\ref{Largebeam}(a) $\langle v\rangle\approx 0.7c_s$ whereas for Fig.~\ref{Narrowbeam} we obtain $\langle v\rangle\approx 3.7c_s$. After the shock is formed in Fig.~\ref{Narrowbeam}, the hole created in the center of the condensate starts to fill in due to the refocusing effect of the trap. Meanwhile, the original outward propagating wavefront has broken into an outer layer of supersonic ripples and an inner primary wavefront moving at the local speed of sound, see also Fig.~\ref{GroundvsLattice}. In the computational frames of Figs.~\ref{Largebeam} and \ref{Narrowbeam} the laser potential is off-centered $3\;\mu$m from the trap/lattice symmetry axis. 

\emph{Sound propagation.---}In order to isolate the effect of vortices on the propagation of density modulations, we simulate perturbations in both the ground state condensate and the rotating vortex lattice and compare the results. In Fig.~\ref{GroundvsLattice} large amplitude density waves are generated using a 5 ms long pulse with $0.15$ mW power and 13 $\mu$m beam waist. The subsequent time-evolution of the condensate density in the two cases can be compared in the sequence of frames for the ground state (lhs) and the vortex lattice state (rhs). The frames for the ground state are taken along the line $y=z=0$, whereas for the lattice the densities in the plane $z=0$ are averaged over circles of radii $r$ in order to integrate out the large density fluctuations due to the vortex cores. To facilitate comparison, the ground state and the vortex lattice state have the same number of atoms. However, the radial trap frequency for the ground state has been weakened by the factor $\sqrt{1-(\Omega/\omega_\perp)^2}$ in order to compensate for the centrifugal effect experienced by the vortex lattice.

In both cases a shock front with a width of the order of the healing length of the condensate is generated, as displayed in the second frames at $t=4$ ms. As this wave reaches the boundary of the driving potential it suddenly slows down to the local speed of sound emitting a collection of high frequency excitations travelling in front of the primary wavefront. These supersonic ripples carry away part of the kinetic energy loaded in the shock-wave. Similar density modulations at the front edge have been recently reported \cite{Kamchatnov2004a,Damski2004a} and were speculated to be a failure of the mean-field theory \cite{Damski2004a}. However, these fringes are clearly present also in the experiments, as is shown in Fig.~\ref{Narrowbeam}, and are a real physical effect. After a quarter of a trap period, the nonrotating condensate forms a density peak in the vicinity of the trap center accompanied by sharp fringes. In the vortex lattice case, however, a centrifugal barrier exists and therefore the center of the trap remains nearly particle free at this stage of the evolution. Similar fringes to those seen in the ground state are also present in the lattice, but they are not visible in the (rhs) of Fig.~\ref{GroundvsLattice} because of the above mentioned averaging. They can, however, be seen in Fig.~\ref{Narrowbeam}.

In the vortex lattice, subjected to the same small amplitude perturbations as the ground state, we find that the average radial speed of sound is equally well described by the equation for $c_s$ evaluated using the chemical potential of the lattice state in place of the one for the ground state. Alternatively, the influence of vortex lattice on the speed of sound may be understood via renormalization of the effective interaction energy by a factor of $b\geq 1$. The value of $b$ (and the chemical potential) depends on the angular velocity of the condensate, resulting in $b\approx 1.16$ in the quantum Hall regime \cite{Baym2004a} and hence to a faster speed of sound with respect to a ground state of the same spatial extent and particle number. This prediction is also in accordance with the velocities measured from our simulations and with the fact that the chemical potential of the lattice state is a factor of $1.1$ larger than that for the corresponding ground state. 

\begin{figure}[!t]
\includegraphics[width=86mm]{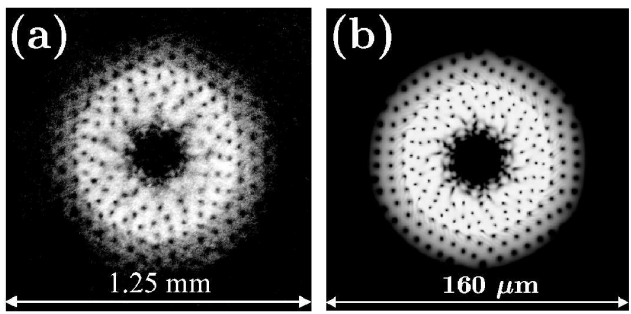}
\caption{Anisotropic sound propagation as observed in an
experiment (a) or in the 3D simulation (b). The duration of the pulse with $P=300\;\mu$W, $d=18 \mu$m is 2.5~ms. Experimental (computational) image taken after 60~ms of anti-trapped expansion (16 ms in-trap evolution).} \label{Anisotropy}
\end{figure}

\emph{Anisotropic sound.---}An interesting feature of sound wave propagation in the vortex lattice is the appearance of an anisotropic wavefront. Initially, the
sound wave reflects the circular symmetry of the laser beam, but each time the wavefront encounters an orbital of vortices, it deforms to
a hexagonal shape following the lattice planes, as depicted in
Fig.~\ref{Anisotropy}, where both the experimental and computational results are shown. On passing through the hexagonal grid of vortices, the wavefront becomes more circular, but is then deformed back to hexagonal
on encountering the next ring of vortices. We note that the supersonic
ripples propagate more isotropically, indicating that the lattice
structure is relatively transparent to these high frequency modulations.

The observed anisotropic sound propagation cannot be explained purely in terms of the static spatial density fluctuations due to the vortex cores. We have shown this by carrying out a simulation for a ground state condensate with a rigid grid of pinning potentials which mimic the density effect of a vortex lattice without phase circulation. In this case the sound waves propagate isotropically without any sign of hexagonal deformations. Thus the development of hexagonal wavefront depends in an essential way on the quantum mechanical current around the vortex cores, the details of which will be discussed elsewhere.

\emph{Discussion.---}We have experimentally studied non-equilibrium dynamics of rapidly rotating Bose-Einstein condensates, triggered by repulsive laser pulses. These experiments are modeled to high accuracy using numerical simulations. Although mean-field theory is known to describe well dilute Bose-Einstein condensates at low temperatures, it is perhaps surprising how accurately it predicts the behavior of rapidly rotating condensates even in the presence of shock waves and after the lattice structure seems to have been completely destroyed. We have identified a sudden increase in the mean-field energy as a signature of the formation of a shock front and obtained a quantitative criterion for it in terms of the kinetic energy absorbed from the beam. The propagation velocity of the peak condensate density is found to be well described by the Bogoliubov speed of sound for both the lattice and the ground state even in the case of large amplitude perturbations. Finally, we have discovered anisotropically propagating sound waves in vortex lattices and showed their inherent relation to the superflow.


\begin{acknowledgments}
Financial support from the Academy of Finland, and the Marsden Fund of New Zealand under Contract No. 02-PVT-004 are acknowledged.
\end{acknowledgments}


\begin{thebibliography}{90}
\bibitem{DonnellyBook}
R.~J. Donnelly, \emph{Quantized Vortices in Helium II} (Cambridge University Press, Cambridge, 1991). 
\bibitem{Madison2000a}
K.~W. Madison \emph{et al.}, Phys. Rev. Lett. {\bf 84}, 806 (2000).
\bibitem{Abo-Shaeer2001a}
J.~R. Abo-Shaeer \emph{et al.}, Science {\bf 292}, 476 (2001).
\bibitem{Haljan2001a}
P.~C. Haljan \emph{et al.}, Phys. Rev. Lett. {\bf 87}, 210403 (2001).
\bibitem{Hodby2002a}
E. Hodby \emph{et al.}, Phys. Rev. Lett. {\bf 88}, 010405 (2002).
\bibitem{Fetterreview}
A.~L. Fetter and A.~A Svidzinsky, J.~Phys.:~Condens.~Matter {\bf 13}, 135 (2001).
\bibitem{Engels2003a}
P. Engels \emph{et al.}, Phys. Rev. Lett. {\bf 90}, 170405  (2003).
\bibitem{Simula2004a}
T.~P. Simula, A.~A. Penckwitt, and R.~J. Ballagh, Phys. Rev. Lett. {\bf 92}, 060401 (2004).
\bibitem{Fetter2003a}
A.~L. Fetter, Phys. Rev. A {\bf 68}, 063617 (2003).
\bibitem{Tkachenko1966a}
V.~K. Tkachenko, Sov. Phys. JETP {\bf 29}, 945 (1969).
\bibitem{Coddington2003a}
I. Coddington \emph{et al.}, Phys. Rev. Lett. {\bf 91}, 100402 (2003).
\bibitem{Baym2004a}
G. Baym, Phys. Rev. Lett. {\bf 91}, 110402 (2003).
\bibitem{Mizushima2004a}
T. Mizushima \emph{et al.}, Phys. Rev. Lett. {\bf 92}, 060407 (2004).
\bibitem{Baksmaty2004a}
L.~O. Baksmaty \emph{et al.}, Phys. Rev. Lett. {\bf 92}, 160405 (2004).
\bibitem{Ho2001a}
T.-L. Ho, Phys. Rev. Lett. {\bf 87}, 060403 (2001).
\bibitem{Schweikhard2004a}
V. Schweikhard \emph{et al.}, Phys. Rev. Lett. {\bf 92}, 040404 (2004).
\bibitem{Baym2004b}
G. Baym and C.~J. Pethick, Phys. Rev. A {\bf 69}, 043619 (2004).
\bibitem{Smith2004a}
N.~L. Smith \emph{et al.}, Phys. Rev. Lett. {\bf 93}, 080406 (2004).
\bibitem{Ketterle1997a}
M.~R. Andrews \emph{et al.}, Phys. Rev. Lett. {\bf 79}, 553 (1997); Erratum: Phys. Rev. Lett. {\bf 80}, 2967 (1998).
\bibitem{Anderson2001a}
B.~P. Anderson \emph{et al.}, Phys. Rev. Lett. {\bf 86}, 2926 (2001).
\bibitem{Dutton2001a}
Z. Dutton \emph{et al.}, Science {\bf 293}, 663 (2001).
\bibitem{Menotti2004a}
C. Menotti \emph{et al.}, Phys. Rev. A {\bf 70}, 023609 (2004).
\bibitem{Coddington2004a}
I. Coddington \emph{et al.}, cond-mat/0405240.
\bibitem{Cozzini2003a}
M. Cozzini and S. Stringari, Phys. Rev. A {\bf 67}, 041602(R) (2003). 
\bibitem{Kamchatnov2004a}
A.~M. Kamchatnov, A. Gammal, and R.~A. Kraenkel, Phys. Rev. A {\bf 69}, 063605 (2004).
\bibitem{Damski2004a}
B. Damski, Phys. Rev. A {\bf 69}, 043610 (2004).

\end{thebibliography}
\end{document}